\documentclass[final,5p,twocolumn,times]{elsarticle}
\usepackage{amssymb}
\usepackage{amsthm}
\usepackage{amsmath}
\usepackage{bm}
\usepackage{color}
\usepackage{float}
\usepackage[dvipsnames]{xcolor}
\usepackage[colorlinks]{hyperref}
\hypersetup{
    citecolor=blue,%
    filecolor=blue,%
    linkcolor=blue,%
    urlcolor=blue
}

\journal{Physica D}

\begin{document}

\begin{frontmatter}

\title{Similarities between characteristics of convective turbulence in confined and extended domains}

\author[label1]{Ambrish Pandey\corref{cor1}}
\ead{ambrish.pandey@nyu.edu}
\cortext[cor1]{Corresponding author}
            
\author[label2]{Dmitry Krasnov}
\ead{dmitry.krasnov@tu-ilmenau.de}
            
\author[label2,label3]{J\"org Schumacher}
\ead{joerg.schumacher@tu-ilmenau.de}

\author[label4]{Ravi Samtaney}

\author[label1,label3,label5]{Katepalli R. Sreenivasan}
\ead{katepalli.sreenivasan@nyu.edu}

\affiliation[label1]{organization={Center for Space Science, New York University Abu Dhabi},
            city={Abu Dhabi},
            postcode={129188}, 
            country={UAE}}
            
\affiliation[label2]{organization={Institute of Thermodynamics and Fluid Mechanics, Technische Universit\"at Ilmenau},            city={Ilmenau},
            postcode={D-98684}, 
            country={Germany}}     

\affiliation[label3]{organization={Tandon School of Engineering, New York University},
            city={New York},
            postcode={NY 11201}, 
            country={USA}}
            
\affiliation[label4]{organization={Division of Physical Science and Engineering, King Abdullah University of Science and Technology},
            city={Thuwal},
            postcode={23955-6900}, 
            country={Saudi Arabia}}
            
\affiliation[label5]{organization={Department of Physics and Courant Institute of Mathematical Sciences, New York University},
            city={New York},
            postcode={NY 11201}, 
            country={USA}}

\begin{abstract}
To understand turbulent convection at very high Rayleigh numbers typical of natural phenomena, computational studies in slender cells are an option if the needed resources have to be optimized within available limits. However, the accompanying horizontal confinement affects some properties of the flow. Here, we explore the characteristics of turbulent fluctuations in the velocity and temperature fields in a cylindrical convection cell of aspect ratio 0.1 by varying the Prandtl number $Pr$ between 0.1 and 200 at a fixed Rayleigh number $Ra = 3 \times 10^{10}$, and find that the fluctuations weaken with increasing $Pr$, quantitatively as in aspect ratio 25. The probability density function (PDF) of temperature fluctuations in the bulk region of the slender cell remains mostly Gaussian, but increasing departures occur as $Pr$ increases beyond unity. We assess the intermittency of the velocity field by computing the PDFs of velocity derivatives and of the kinetic energy dissipation rate, and find increasing intermittency as $Pr$ decreases. In the bulk region of convection, a common result applicable to the slender cell, large aspect ratio cells, as well as in 2D convection, is that the turbulent Prandtl number decreases as $Pr^{-1/3}$.
\end{abstract}


\end{frontmatter}

\section{Introduction} 
\label{sec:intro}

Many turbulent flows in nature are driven by thermal convection. The phenomenon of tectonic plate and the generation and sustenance of the geomagnetic field are manifestations of thermal convection in the Earth's mantle and outer core, respectively~\citep{Schubert:book2001, Turcotte:book2002}. Sustained magnetic field and its stochastic excursions in the Sun may well be connected to interior convection; for example, see \citep{Schumacher:RMP2020, Pandey:PRF2021}. Rayleigh-B\'enard convection (RBC), where a fluid layer is heated from the bottom and cooled from the top, is a paradigm of such flows~\citep{Sreenivasan:1998, Chilla:EPJE2012, Verma:book2018}. The Prandtl number $Pr$, which is the ratio of the heat and momentum diffusion time scales in the fluid, and a crucial governing parameter of RBC, varies immensely in natural flows. For instance, it changes dramatically in the Earth's interior from $10^{-1}$ in the outer core~\citep{Calkins:EPSL2012} to $10^{24}$ in the mantle~\citep{Schubert:book2001}. The Rayleigh number $Ra$, signifying the ratio of the driving buoyancy force to dissipative forces due to viscosity and thermal diffusivity, and the aspect ratio $\Gamma$, which is the ratio of horizontal to the vertical dimension of fluid apparatus, are two other governing parameters of RBC. The Nusselt number $Nu$ quantifies the heat transport enhancement due to convective motion and the Reynolds number $Re$ indicates the vigor of turbulence; both response parameters $Nu$ and $Re$ are functions of the control parameters of RBC, namely $Ra$, $Pr$ and $\Gamma$; see, for example, \citep{Chilla:EPJE2012, Ahlers:RMP2009, Verma:NJP2017}. In this paper, we compare the properties of velocity and temperature fluctuations in confined convection with those in horizontally extended domains for a wide range of $Pr$.

For a given Rayleigh number, the computational power required to explore convective flows varies nominally as $\Gamma^2$. For this reason, it is tempting to use slender convection cells to achieve very high $Ra$. \citet{Iyer:PNAS2020} studied the scaling features of global heat and momentum transports in a cylindrical cell at aspect ratio $\Gamma=D/H=0.1$ for a fixed $Pr = 1$ and $Ra$ up to $10^{15}$, and found that the power law scaling of $Nu(Ra)$ is very close to $Nu\sim Ra^{1/3}$, in agreement with the marginally stable boundary layer model of Malkus \cite{Malkus:PRSL1954a}; see also the commentary by Doering \cite{Doering:PNAS2020}. However, the rate of increase of $Re$ with $Ra$ was found to be lower than that in wider domains~\citep{Iyer:PNAS2020}. \citet{Pandey:EPL2021} explored the $Pr$-dependence of $Nu$ and $Re$ in the same $\Gamma = 0.1$ convection cell and observed that, while heat transport was lower compared to those in wider cells when $Pr$ and $Ra$ are small, they approach those in wider domains when $Pr \geq 1$ and $Ra \geq 10^{10}$.

In the present work, we inspect the $Pr$-dependence of several quantities in the slender cell, such as the probability density functions (PDF) of the temperature and velocity fluctuations and of the corresponding dissipation rates, and find that they are similar to those in wider domains. The turbulent Prandtl number $Pr_t$ in the slender cell also exhibits a very similar $Pr$-dependence to that in $\Gamma \geq 1$ convection cells~\citep{Pandey:PRF2021}. The turbulent (or effective) Prandtl number $Pr_t$, which is the ratio of the effective viscosity $\nu_t$ and the effective thermal diffusivity $\kappa_t$ in turbulent flows, is pivotal in modelling engineering and atmospheric turbulence~\citep{Li:AR2019}. Because of highly efficient stirring, turbulent flows are endowed with far larger turbulent diffusivities $\nu_t$ and $\kappa_t$ than molecular diffusivities $\nu$ and $\kappa$. $Pr_t \approx 1$ in moderate-$Pr$ flows~\citep{Abe:IJHMT2019}, as expected from the Reynolds analogy that the same eddies transport heat as well as momentum ~\citep{Li:AR2019}. However, $Pr_t > 1$ have been consistently found in low-Prandtl-number fluids~\citep{Abe:IJHMT2019, Bricteux:NED2012, Reynolds:IJHMT1975}. Recently, \citet{Pandey:PRF2021} estimated $Pr_t$ in RBC for a wide range of $Pr$ and found that $Pr_t \sim Pr^{-1/3}$ when the molecular Prandtl number decreased from 13 to $10^{-4}$ for a constant Grashof number, $Gr = Ra/Pr = 10^9$. Here, we find the same trends with and without confinement, and $Pr_t$ remains nearly unaffected by the horizontal confinement as long as the thermal forcing is strong enough.

To be specific, we study thermal convection in the following cases: a cylinder whose diameter is one-tenth of its height at $Ra = 3 \times 10^{10}$, for a wide range of Prandtl numbers at 12 different values between $0.1 \leq Pr \leq 200$. We make a systematic study of the dependence of the statistics on the Prandtl number and relate the results to recent studies at larger aspect ratios, paying particular attention to the demanding resolution of time and length scales at low Prandtl numbers \citep{Schumacher:PNAS2015}. The dominant force balance in low-$Pr$ convection is between the inertial and pressure gradient forces, which leads to vigorous turbulence compared to those at moderate and high $Pr$~\citep{Pandey:POF2016}. Due to increasing availability of computational resources, low-$Pr$ regime of convection, of great interest in geophysical and atmospheric flows, can now be computed~\citep{Schumacher:PNAS2015, Scheel:PRF2017, Pandey:Nature2018, Zwirner:JFM2020}. The velocity field in such flows exhibits a highly intermittent character, whereas the temperature field remains diffusive. We quantify the intermittency by computing the flatness factor of the velocity and temperature fluctuations and of their derivatives, and observe that the temperature (velocity) field becomes increasingly intermittent as $Pr$ increases (decreases). This feature is consistent with the characteristics of fluctuations for $\Gamma \geq 0.5$. 

The remainder of the paper is organized as follows. We present the parameters of our simulations in Sec.~\ref{sec:numerical}. The flow structure in the slender cell is described in Sec.~\ref{sec:flow_str}. The properties of the temperature and velocity fluctuations are quantified in Secs.~\ref{sec:T_fluct} and~\ref{sec:v_fluct}, respectively. The variation of the turbulent Prandtl number with the molecular Prandtl number is presented in Sec.~\ref{sec:Prt}, and the main findings are summarized in Sec.~\ref{sec:conclu}.

\section{Governing equations and numerical details}
\label{sec:numerical}

We perform direct numerical simulations (DNS) of Oberbeck-Boussinesq convection by solving the following non-dimensional equations
\begin{align}
\frac{D u_i}{Dt}  & =   -\frac{\partial p}{\partial x_i} + T \delta_{iz} + \sqrt{\frac{Pr}{Ra}} \,  \frac{\partial^2 u_i}{\partial x_j^2}, \label{eq:u} \\
\frac{D T}{Dt}  & =  \frac{1}{\sqrt{PrRa}} \, \frac{\partial^2 T}{\partial x_j^2}, \label{eq:T} \\
\frac{\partial u_j}{\partial x_j}  & =  0, \label{eq:m}
\end{align}
where $u_i$ (with $i \equiv x,y,z$), $p$, and $T$ are the velocity, pressure, and temperature fields, respectively. The operator $D/Dt = \partial/\partial t + u_j \partial /\partial x_j$ is the substantial derivative. These equations are non-dimensionalized using the depth $H$, the temperature difference $\Delta T$ between the horizontal plates, the free-fall velocity $u_f = \sqrt{\alpha g \Delta T H}$, and the free-fall time $t_f = \sqrt{H/(\alpha g \Delta T)}$ as the length, temperature, velocity, and time scales, respectively. The non-dimensional governing parameters are defined as $Ra = \alpha g \Delta T H^3/(\nu \kappa)$ and $Pr = \nu/\kappa$, where $\alpha, \nu, \kappa$ are respectively the isobaric coefficient of thermal expansion, the kinematic viscosity, and the thermal diffusivity of the working fluid, and $g$ being the acceleration due to gravity. Note that in experimental investigations of liquid metal convection in slender cells~\citep{Frick:EPL2015, Mamykin:MHD2015}, the cell diameter is also used as the relevant length scale to define the Rayleigh number. This rescaling will decrease the values of $Ra$ by a factor of $10^3$ for the present $\Gamma$. Depending on the length scale used in the definition of the Rayleigh number, different prefactors in transport scaling laws might result. Our focus is on the dependence with respect to $Pr$.

Simulations are performed using the solver {\sc Nek5000}, which is based on the spectral element method~\citep{Fischer:JCP1997}. The simulation domain consists of $N_e$ elements, with each element further discretized using Lagrangian interpolation polynomials of order $N$, resulting in $N_e N^3$ mesh cells in the entire flow domain. For the temperature field, we employ the isothermal and adiabatic conditions on the horizontal plates and the sidewall, respectively. The velocity field satisfies no-slip condition on all boundaries. Increased density of mesh cells is used in thermal and viscous boundary layers (BLs) to adequately resolve them.

For cells with aspect ratio 0.1, we perform DNS for $Ra = 3 \times 10^{10}$, and $0.1 \leq Pr \leq 200$. Simulations are started from the conduction solution with random perturbations, and are continued well into the statistically steady state for a total time $t_\mathrm{sim}$. The simulation data from \citet{Pandey:EPL2021} have been extended for longer times to substantiate converged statistics. We resolve both the Kolmogorov length scale $\eta_u$ and the Batchelor length scale $\eta_B$, the finest scales of velocity and temperature, respectively. The local Kolmogorov scale is estimated as
\begin{equation}
\eta_u(x_i) = \left( \frac{\nu^3}{\varepsilon_u(x_i)} \right)^{1/4},
\end{equation}
where 
\begin{equation}
\varepsilon_u(x_i) = 2 \nu S_{ij} S_{ij} \label{eq:eps_u}
\end{equation}
is the kinetic energy dissipation rate per unit mass and
\begin{equation}
S_{ij} = \frac{1}{2} \left( \frac{\partial u_i}{\partial x_j} + \frac{\partial u_j}{\partial x_i} \right)
\end{equation}
is the strain rate tensor. The Batchelor scale $\eta_B = \eta_u/\sqrt{Pr}$ is finer than the Kolmogorov scale when $Pr > 1$. We compute the depth-dependent Kolmogorov and Batchelor scales using the area- and time-averaged energy dissipation rate, and compare them with the vertical grid spacing $\Delta_z(z)$ for each simulation. The maximum values of $\Delta_z(z)/\eta_u(z)$ for $Pr \leq 1$ and $\Delta_z(z)/\eta_B(z)$ for $Pr > 1$ are listed in Table~\ref{tab_ra3e10}. These ratios are of the order unity, which indicates that the spatial resolution is adequate in all simulations~\citep{Scheel:NJP2013}.

\begin{table}
\caption{DNS parameters for $Ra = 3 \times 10^{10}$ in a slender cylinder of $\Gamma = 0.1$. The number of elements $N_e = 192,000$ for all except that at $Pr = 0.1$, for which $N_e = 537,600$. Both element numbers have been used in \cite{Iyer:PNAS2020} for $Ra\le 10^{11}$ and $Ra=10^{12}$ (at fixed $Pr=1$). $N$ is the order of the Lagrangian interpolation polynomials; $Nu$ and $Re$ are the globally-averaged heat and momentum transports, respectively; $t_\mathrm{sim}$ is the total duration of simulations in units of the free-fall time in statistically steady state; $\Delta_z/\eta$ is the maximum value of the ratio of vertical grid spacing to the local Kolmogorov (for $Pr \leq 1$) or Batchelor (for $Pr > 1$) scale in the flow. The error bars in $Nu$ and $Re$ represent the difference between the mean values computed over the two halves of the datasets. In particular, all the simulations have been integrated for longer periods than in \citet{Pandey:EPL2021} to confirm converged statistics.}
\label{tab_ra3e10}
\begin{center}
\begin{tabular}{lccccc}
\hline
$Pr$ & $N$ & $Nu$ & $Re$ & $t_\mathrm{sim} \, (t_f)$ & $\Delta_z/\eta$  \\
\hline
0.1 & 11 & $104.2 \pm 2.5$ & $44472 \pm 150$ & 44 & 1.49 \\
0.2 & 11 & $112.0 \pm 15$ & $27634 \pm 212$ & 90 & 1.45 \\
0.35 & 9 & $131.4 \pm 8.0$ & $19688 \pm 76$ & 132 & 1.38 \\
0.5 & 7 & $134.3 \pm 9.2$ & $15173 \pm 64$ & 191 & 1.48 \\
0.7 & 7 & $146.2 \pm 1.7$ & $12248 \pm 9$ & 165 & 1.26 \\
1 & 7 & $150.0 \pm 1.2$ & $9411 \pm 2$ & 213 & 1.08 \\
2 & 7 & $162.4 \pm 7.7$ & $5673 \pm 10$ & 270 & 1.09 \\
4.38 & 7 & $177.2 \pm 0.7$ & $3230 \pm 3$ & 282 & 1.12 \\
7 & 7 & $181.8 \pm 2.7$ & $2284 \pm 1$ & 254 & 1.12 \\
20 & 7 & $181.8 \pm 7.6$ & $965 \pm 2$ & 478 & 1.12 \\
100 & 7 & $176.6 \pm 0.4$ & $205 \pm 1$ & 454 & 1.12 \\
200 & 7 & $179.3 \pm 0.4$ & $103 \pm 1$ & 858 & 1.12 \\
\hline
\end{tabular}
\end{center}
\end{table}

\begin{figure}
\begin{center}
\includegraphics[width=0.5\textwidth]{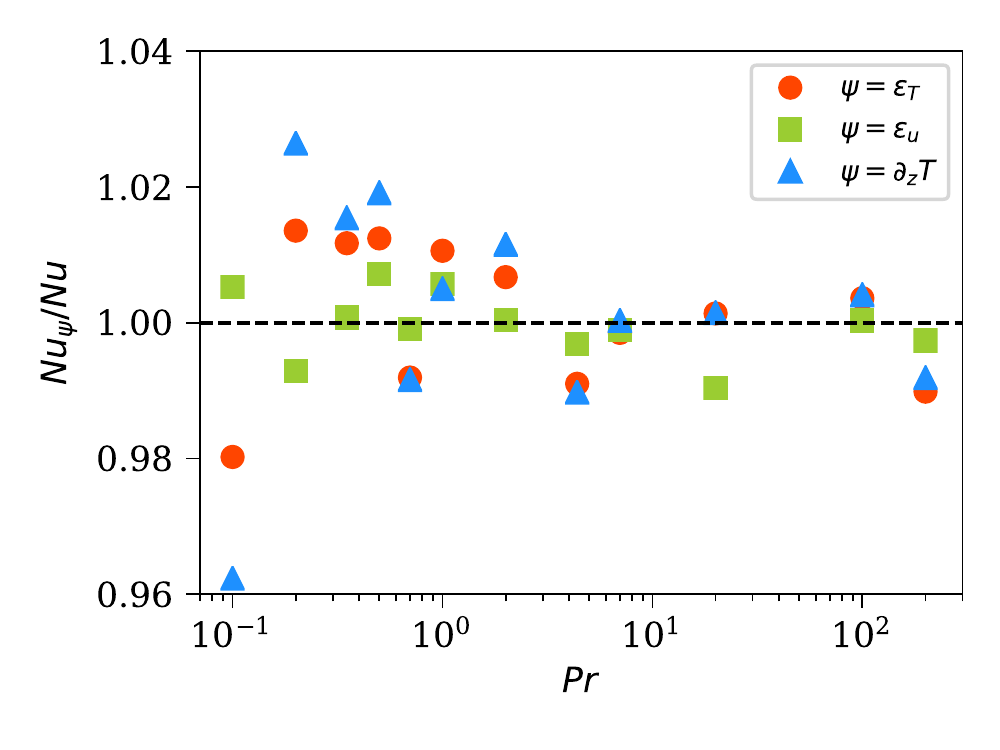}
\caption{Normalized Nusselt numbers $Nu_{\varepsilon_T}/Nu$, $Nu_{\varepsilon_u}/Nu$, and $Nu_{\partial_z T}/Nu$ for all the simulations. The concurrence of heat transport from all the methods to within 4\% indicates the adequacy of spatial and temporal resolutions of the simulations.}
\label{fig:Nu_four}
\end{center}
\end{figure}

The adequacy of spatial and temporal resolutions can be further ensured by computing the turbulent heat flux using different methods. A significant enhancement of heat transport over the hydrostatic case occurs due to the convective motion of the fluid. The Nusselt number $Nu$ quantifies this enhancement and is computed as
\begin{equation}
Nu = 1 + \sqrt{Ra Pr} \, \langle u_z T \rangle_{V,t} \, , \label{eq:Nu_uzT}
\end{equation}
where $\langle \cdot \rangle_{V,t}$ denotes averaging over the entire simulation domain and integration time. The global heat transport can also be estimated from the viscous and thermal dissipation rates using the exact relations~\citep{Howard:ARFM1972, Shraiman:PRA1990} as
\begin{align}
Nu_{\varepsilon_u} & =  1+ \sqrt{RaPr} \, \langle \varepsilon_u \rangle_{V,t} \, , \label{eq:Nu_epsv} \\
Nu_{\varepsilon_T} & =  \sqrt{RaPr} \, \langle \varepsilon_T \rangle_{V,t} \, ,	\label{eq:Nu_epst}
\end{align}
where the thermal dissipation rate $\varepsilon_T$ is computed as
\begin{equation}
\varepsilon_T(x_i) = \kappa \left( \frac{\partial T}{\partial x_i} \right)^2 . \label{eq:eps_T}
\end{equation}
Further, the time- and area-averaged heat flux at each depth should match those computed using the above methods. We estimate the heat flux at the bottom and top plates as
\begin{equation}
Nu_{\partial_z T} = -\left\langle \left( \frac{\partial T}{\partial z}\right)_{z=0,H} \right\rangle_{A,t} \, , \label{eq:Nu_plate}
\end{equation}
where $\langle \cdot \rangle_{A,t}$ stands for averaging over both the area and time. The normalized Nusselt numbers $Nu_{\varepsilon_T}/Nu$, $Nu_{\varepsilon_u}/Nu$, and $Nu_{\partial_z T}/Nu$, shown in Fig.~\ref{fig:Nu_four}, confirm that the heat fluxes from all the methods converge to within 4\%, reassuring that all simulations are adequately resolved.

\section{Flow structure in slender domain}
\label{sec:flow_str}

\begin{figure*}
\begin{center}
\includegraphics[width=0.75\textwidth]{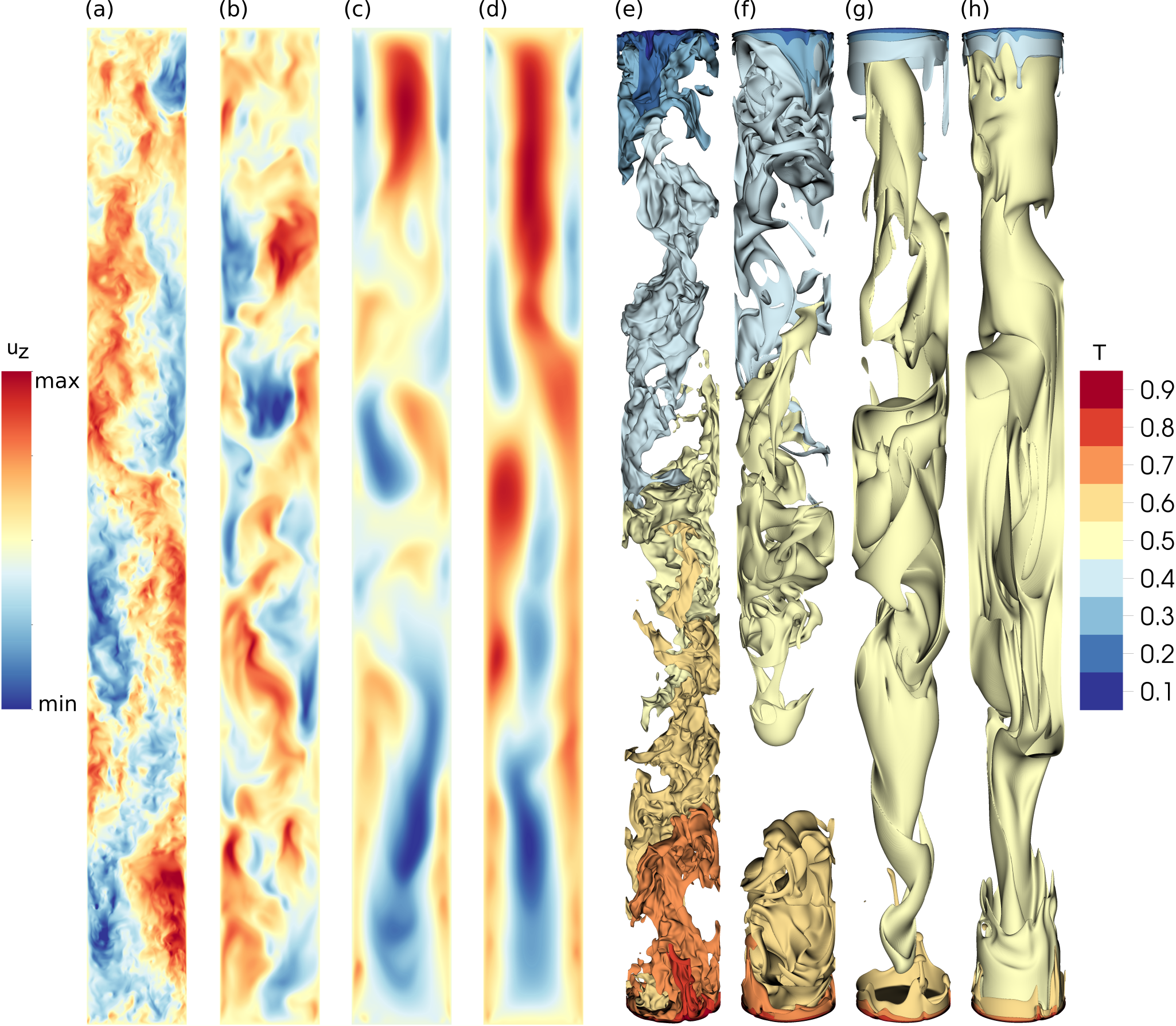}
\caption{Vertical velocity contours in a vertical plane (a--d) and temperature isosurfaces (e--h) for $Pr = 0.2$ (a,e), $Pr = 2$ (b,f), $Pr = 20$ (c,g), $Pr = 200$ (d,h). Velocity structures become increasingly intermittent with decreasing $Pr$ and the bulk region becomes increasingly isothermal with increasing $Pr$.}
\label{fig:flow_ver}
\end{center}
\end{figure*}

\begin{figure*}
\begin{center}
\includegraphics[width=0.8\textwidth]{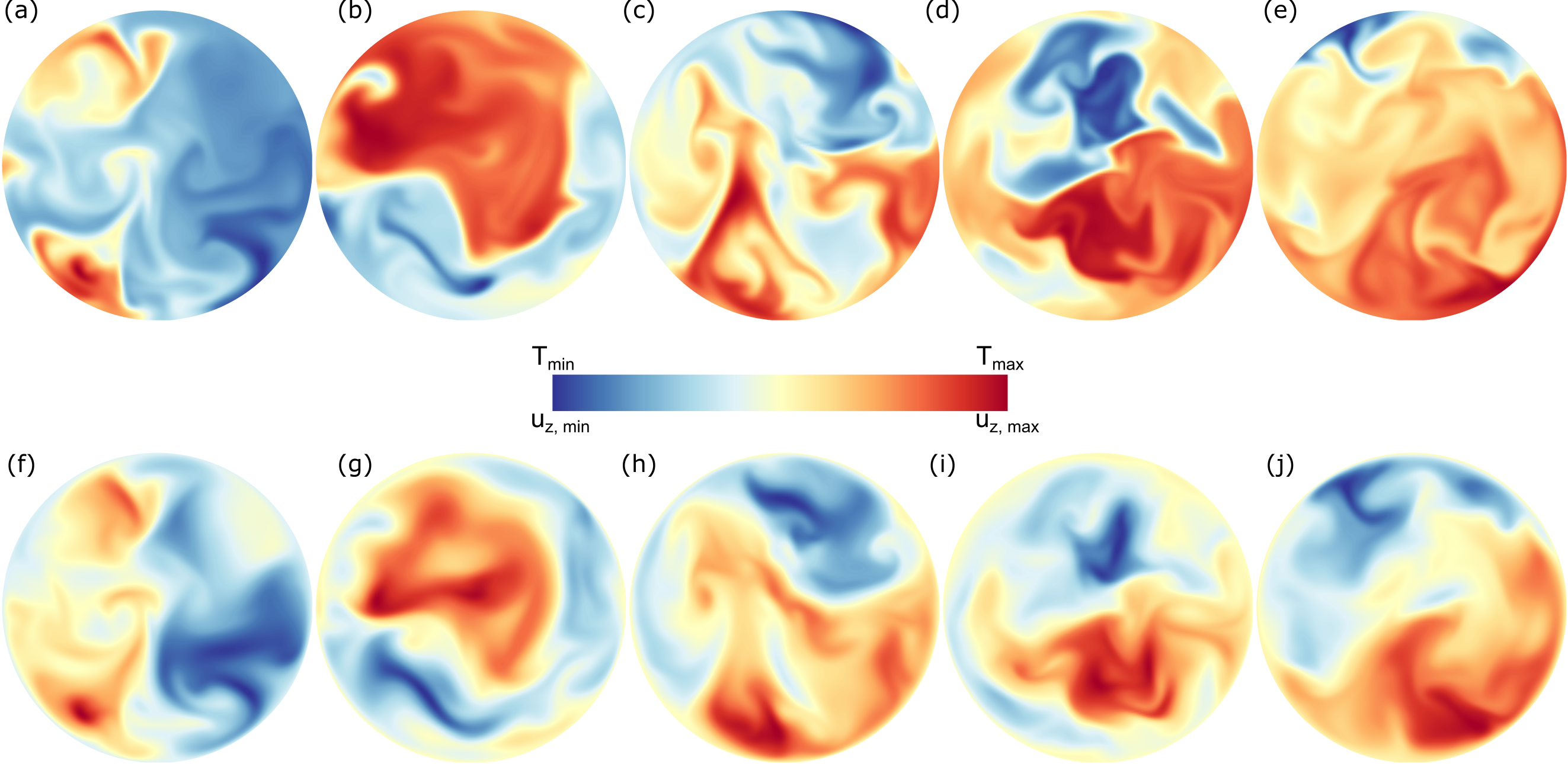}
\caption{Temperature field $T(x,y)$ (top panels) and vertical velocity field $u_z(x,y)$ (bottom panels) for $Pr = 1$: (a,f) $z = 0.10H$, (b,g) $z = 0.25H$, (c,h) $z = 0.50H$, (d,i) $z = 0.75H$, (e,j) $z = 0.90H$. The locations of the hot up-wellings and cold down-wellings shift as we move in the vertical direction (from left to right), consistent with the helical flow structure.}
\label{fig:flow_hor}
\end{center}
\end{figure*}

A single convection roll occupying the entire domain is observed for $\Gamma \approx 1$ ~\citep{Scheel:PRF2017, Niemela:JFM2003}, whereas complex patterns consisting of an array of rolls are observed in horizontally-extended domains~\citep{Pandey:Nature2018, Fonda:PNAS2019}. In slender cells, however, multiple vertically-stacked convection rolls are observed \citep{ Niemela:JFM2003, Verzicco:JFM2003}, and are attributed to the elliptical instability~\citep{Zwirner:PRL2020}. The flow organization in the slender cell for different Prandtl numbers is demonstrated in Fig.~\ref{fig:flow_ver} which plots the vertical velocity in a vertical plane, as well as temperature isosurfaces.  The locations of the up- and down-wellings are exchanged multiple times in the flow at $Pr = 0.2$ (Fig.~\ref{fig:flow_ver}(a)), which suggests the presence of helical flow structure similar to a barber pole~\citep{Iyer:PNAS2020}. Figure~\ref{fig:flow_ver} further shows that increasingly finer structures appear in the velocity field as $Pr$ decreases, which is due to increasing Reynolds numbers. The instantaneous temperature isosurfaces reveal that a significant temperature variation exists even in the bulk region for $Pr = 0.2$, which becomes weaker as $Pr$ increases. The elongated isosurfaces corresponding to the mean temperature $\Delta T/2$ in Fig.~\ref{fig:flow_ver}(g, h) indicate that the bulk of the flow in the slender cell is nearly isothermal for large Prandtl numbers~\citep{Pandey:EPL2021}.

The helical structure observed in Fig.~\ref{fig:flow_ver}(a) can be probed further by examining the vertical velocity and temperature slices in different horizontal planes. Figure~\ref{fig:flow_hor} shows $T(x,y)$ and $u_z(x,y)$ for $Pr = 1$ at five different depths in the slender cell and reaffirms that the hot up-welling and cold down-welling regions shift as the top plate is approached, in a manner that is consistent with the helical structure~\citep{Iyer:PNAS2020}. Further, the correlation between velocity and temperature structures remains strong, leading to a strong convective heat flux in the slender cell, despite weaker turbulence~\citep{Pandey:EPL2021}. Figure~\ref{fig:flow_hor} also shows that the smallest scales in both the temperature and velocity fields are similar (Kolmogorov and Batchelor scales are the same for $Pr = 1$).

\begin{figure}
\begin{center}
\includegraphics[width=0.5\textwidth]{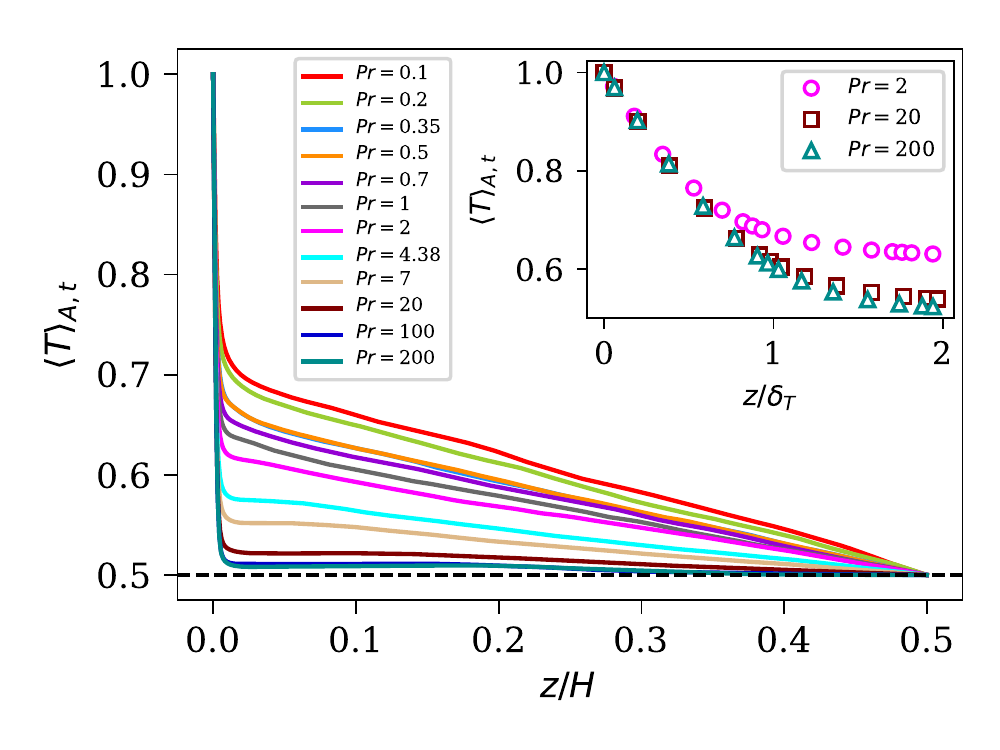}
\caption{Mean temperature profiles averaged over the top and bottom halves of the domain exhibit that the mean gradient in the bulk region of the slender cell decreases with increasing $Pr$ and approaches the limit of a well-mixed bulk with no mean gradient, as observed in large-aspect-ratio convection. Inset reveals that the temperature variation in the near-wall region is linear and the number of nodes within the thermal BL region is adequate to capture the strong temperature gradients.}
\label{fig:T_z}
\end{center}
\end{figure}

To quantify the vertical variation of temperature as exhibited in Fig.~\ref{fig:flow_ver}(e-f), we plot the mean temperature profiles $\langle T \rangle_{A,t}(z)$ in Fig.~\ref{fig:T_z}. Exploiting the up-down symmetry of Boussinesq convection, we average the profiles over the top and bottom halves of the domain to obtain better statistics. Figure~\ref{fig:T_z} shows that, in addition to a rapid variation in the thermal boundary layer (BL) region, a significant temperature variation occurs in the bulk region up to $Pr = 7$. However, for $Pr \geq 20$, the temperature gradient in the bulk region becomes very close to that ($\approx 0$) observed in wider convection cells. The inset of Fig.~\ref{fig:T_z} shows the temperature variation in the vicinity of the bottom plate for three large-$Pr$ simulations. We observe that the temperature varies linearly in the near-wall region. The inset also highlights that there are 8-9 mesh cells within the thermal BL region for the largest $Pr$ cases, thus confirming the adequacy of resolution for capturing rapid variations of temperature near the bottom and top plates~\citep{Scheel:NJP2013}.


\section{Characteristics of temperature fluctuations}
\label{sec:T_fluct}

\begin{figure}
\begin{center}
\includegraphics[width=0.5\textwidth]{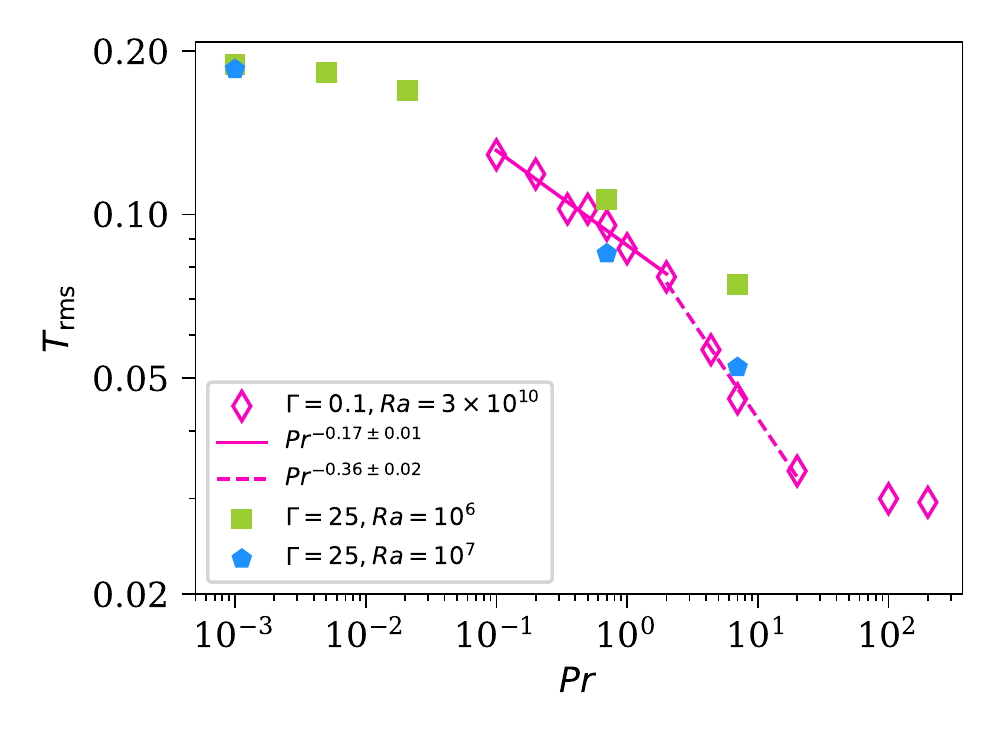}
\caption{Root-mean-square fluctuation about the global mean temperature decreases with increasing $Pr$ in both the slender and extended convection domains, with nearly similar magnitude in the overlapping range of $Pr$. The solid and the dashed lines are the best fits to the slender cell data.}
\label{fig:T_rms}
\end{center}
\end{figure}

The average temperature $\langle T \rangle_{V,t}$ in Oberbeck-Boussinesq convection remains $\Delta T/2$. However, the strength of fluctuations depends on the governing parameters. We compute the root-mean-square (rms) temperature fluctuation about the global mean as 
\begin{equation}
T_\mathrm{rms} = \left( \left\langle (T - \langle T \rangle_{V,t})^2 \right \rangle_{V,t} \right)^{1/2} =  \left( \langle T^2 \rangle_{V,t} - \langle T \rangle^2_{V,t} \right)^{1/2}
\end{equation}
and show that $T_\mathrm{rms}$ decreases with increasing $Pr$ up to $Pr \approx 100$ and tends to saturate thereafter (Fig.~\ref{fig:T_rms}). \citet{Daya:PRE2002} measured the temperature fluctuations at the center of a cylindrical domain with $\Gamma \approx 0.8$ and found that the rms fluctuation decreases with both $Ra$ and $Pr$.  Figure~\ref{fig:T_rms} shows that $T_\mathrm{rms}$ in the slender cell decreases gradually for $Pr \leq 2$ and rapidly for larger $Pr$ up to $Pr = 20$. The best fit for $0.1 \leq Pr \leq 2$ yields $T_\mathrm{rms} = (0.088 \pm 0.001) Pr^{-0.17 \pm 0.01}$, and $T_\mathrm{rms} = (0.096 \pm 0.002) Pr^{-0.36 \pm 0.02}$ for $2 \leq Pr \leq 20$. The latter power law exponent of $-0.36$ is close to $-0.38 \pm 0.04$ reported by \citet{Daya:PRE2002} in a narrow Prandtl number range $3 \leq Pr \leq 12.3$. We also compute $T_\mathrm{rms}$ for $Ra = 10^{10}$ in the slender cell~\citep{Pandey:EPL2021} and obtain that it is larger than the corresponding values at $Ra = 3 \times 10^{10}$ (not shown), again consistent with convection in wider cells~\citep{Daya:PRE2002, Niemela:Nature2000, Zhou:JFM2013}. We find that the rms fluctuation of the deviation of temperature from the diffusive equilibrium profile increases 
weakly with $Pr$ (not shown), which is consistent with the profiles exhibited in Fig.~\ref{fig:T_z}; see also \citet{Pandey:PRE2014, Pandey:POF2016}.

The effect of horizontal confinement on temperature fluctuations is examined by plotting $T_\mathrm{rms}$ in a $\Gamma = 25$ cube in Fig.~\ref{fig:T_rms} (\citet{Pandey:JFM2022}). $T_\mathrm{rms}$ for $Ra = 10^7$ in the extended domain (blue pentagons) is comparable to that in the slender case. Also, $T_\mathrm{rms}$ for $Ra = 10^6$ are a bit larger, consistent with the decreasing trend with $Ra$~\citep{Daya:PRE2002, Niemela:Nature2000, Zhou:JFM2013}. The thermal plumes, which are hotter or colder than the ambient fluid, are primarily responsible for temperature fluctuations in the flow. The increase of $T_\mathrm{rms}$ with decreasing $Pr$ could therefore be associated with increasing coarseness of the plumes~\citep{Pandey:JFM2022}. Note that the critical Rayleigh number $Ra_c$ for the onset of convection in the slender cell is approximately $1.1 \times 10^7$~\citep{Pandey:EPL2021}, which is three orders of magnitude larger than that for $\Gamma = 25$~\citep{Shishkina:PRF2021}. The supercritical Rayleigh number for the slender domain is $Ra/Ra_c \approx 2700$ and for the data at $Ra = 10^7$ in the extended cell is $Ra/Ra_c = 5500$. Thus, the similarity of $T_\mathrm{rms}$ in both domains seems to indicate that the properties of temperature fluctuations are unaffected by aspect ratio as long as the relative strengths of the thermal forcing are comparable.

\begin{figure}[t]
\begin{center}
\includegraphics[width=0.5\textwidth]{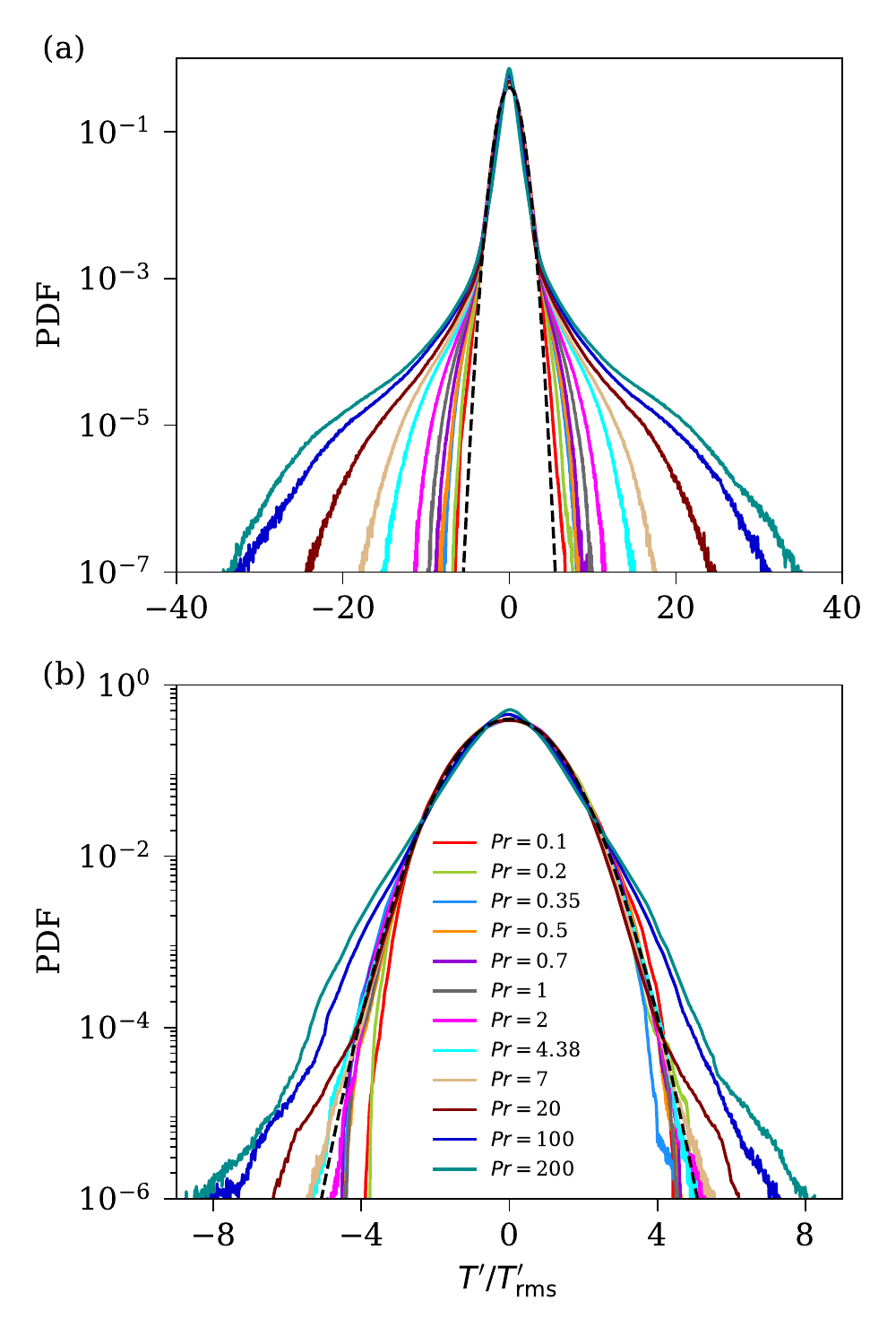}
\caption{Probability density function (PDF) of the normalized temperature fluctuations $T^\prime/T^\prime_\mathrm{rms}$ over (a) the entire domain and (b) only the bulk region. The PDFs are nearly Gaussian for low Prandtl numbers but departures from Gaussianity increases as $Pr$ increases.}
\label{fig:pdf_Tp}
\end{center}
\end{figure}

To further characterize turbulent fluctuations, we decompose the velocity and temperature fields as
\begin{eqnarray}
u_i(x_k,t) & = & U_i(x_k) + u^\prime_i(x_k,t), \\
T(x_k,t) & = & \Theta(x_k) + T^\prime(x_k,t),
\end{eqnarray}
where ($U_i, \Theta$) are the mean components and ($u^\prime_i, T^\prime$) are the fluctuating components of the
velocity and temperature fields, respectively. Here, again $i,k=x,y,z$. The first moment of $T^\prime$, i.e., the volume- and time-averaged temperature fluctuation, vanishes. Furthermore, the variation of the rms fluctuation $T^\prime_\mathrm{rms}$ is approximately similar to $T_\mathrm{rms}(Pr)$, although the magnitude of the former is smaller by nearly a factor of three.

As a further characterization of temperature fluctuations, we plot in Fig.~\ref{fig:pdf_Tp}(a) the probability density function (PDF) of $T^\prime$ over the entire domain. We find that the core of all the PDFs closely follows a Gaussian distribution for $|T^{\prime}/T^{\prime}_{\rm rms}|\lesssim 4$. However, the tails (large amplitude fluctuations) are Gaussian only for low Prandtl numbers up to $Pr \approx 0.35$. Departure from Gaussianity in the tails is evident for $Pr \geq 1$. Figure~\ref{fig:pdf_Tp}(b) shows the PDFs of $T^\prime$ over the bulk region, i.e., for $0.2H \leq z \leq 0.8H$; the distribution in the bulk region is very close to Gaussian for $Pr \leq 7$. The non-Gaussian fat tails start to appear only for $Pr \geq 20$. The disappearance of large amplitude events from the bulk PDFs indicates that thermal plumes, which are the primary candidates for fat tails, undergo a strong mixing in the bulk region. The variation with $Pr$ in the bulk region of the slender cell is similar to that in extended domains, as reported in Fig.~5 of \citet{Schumacher:PRE2018}.

\begin{figure}
\begin{center}
\includegraphics[width=0.5\textwidth]{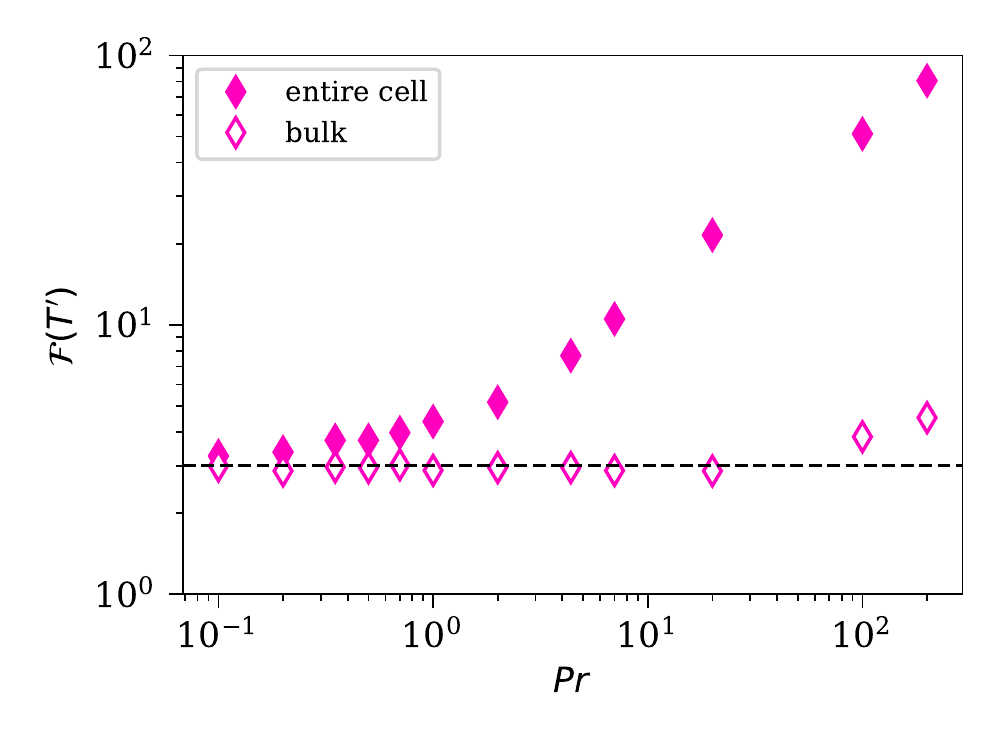}
\caption{Normalized fourth-order moment of the temperature fluctuations $T^\prime$ in the entire cell (filled symbols) and in the bulk region (open symbols). Dashed line is the Gaussian value. Strong departures from the Gaussian distribution in high-$Pr$ convection are evident.}
\label{fig:kurtosis}
\end{center}
\end{figure}

Even though the PDFs depart from Gaussianity with increasing $Pr$, they remain closely symmetric (Fig.~\ref{fig:pdf_Tp}). The normalized fourth-order moment of $T^\prime$, defined as
\begin{equation}
\mathcal{F}(T^\prime) = \frac{\langle T^{\prime 4} \rangle}{\langle T^{\prime 2} \rangle^2},
\end{equation}
is a measure of intermittency in the distribution~\citep{Sreenivasan:ARFM1997, Pandey:PRE2018}.  Figure~\ref{fig:kurtosis} shows that the flatness in the bulk region (open diamonds) remains  close to that for the Gaussian distribution (=3) for all but the highest two Prandtl numbers. However, the flatness over the entire domain (filled diamonds) departs significantly from 3 for $Pr > 1$, consistent with earlier discussion that $T^\prime$ in the bulk follows the Gaussian distribution, while the plumes in the thermal BL lead to large amplitude events and stretched tails in the PDFs of $T^\prime$. Due to very low thermal diffusivity in high-$Pr$ convection, the plumes preserve their identity longer and significantly contribute to large fluctuation characteristics~\citep{Schumacher:PRE2018, Silano:JFM2010, Pandey:PRE2014}.


\section{Characteristics of velocity fluctuations}
\label{sec:v_fluct}

\begin{figure}
\begin{center}
\includegraphics[width=0.5\textwidth]{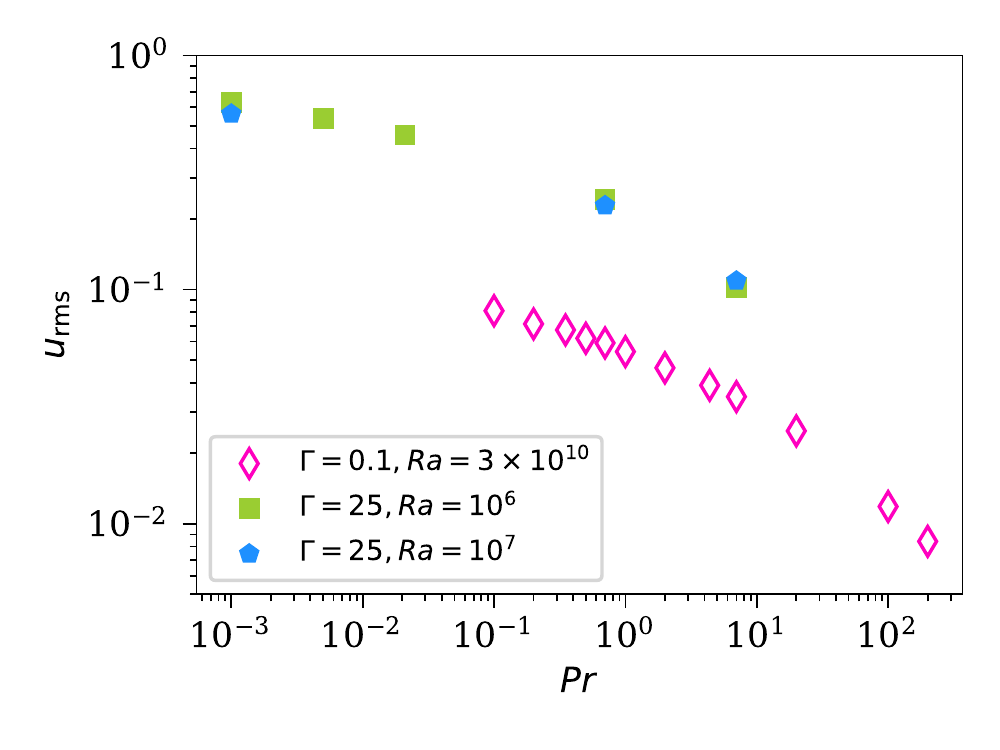}
\caption{Root-mean-square velocity decreases with increasing $Pr$. Filled symbols are from the simulations of \citet{Pandey:JFM2022} in the extended convection cell. Velocity fluctuations in the slender domain are weaker than those in the extended domain, but the variation with $Pr$ is similar in both configurations.}
\label{fig:urms}
\end{center}
\end{figure}

The rms velocity,
\begin{equation}
u_\mathrm{rms} = \sqrt{\langle u_i^2 \rangle_{V,t}} = \sqrt{\langle u_x^2 + u_y^2 + u_z^2 \rangle_{V,t}},
\end{equation}
is plotted as a function of $Pr$ in Fig.~\ref{fig:urms}. We observe that it decreases with increasing $Pr$, owing to the disproportionate increase of the viscous effects. Figure~\ref{fig:urms} also shows $u_\mathrm{rms}$ obtained in the extended domain at $\Gamma=25$ (\citet{Pandey:JFM2022}). It is seen that $u_\mathrm{rms}$ in the slender cell is nearly four times smaller, confirming the earlier results of~\citet{Iyer:PNAS2020} for different Prandtl numbers. Further, the functional form of $u_\mathrm{rms}(Pr)$ is similar in both configurations; a gradual decrease with $Pr$ in the turbulent regime is followed by a steeper decrease in the viscous regime~\citep{Pandey:EPL2021, Pandey:POF2016}.

We have computed (but not shown) the PDFs of the velocity components in the slender cell and find that the vertical velocity fluctuations are Gaussian. However, the PDFs of the horizontal velocity fluctuations show increasing departure from Gaussianity with increasing $Pr$. In quantitative agreement, the flatness of the vertical velocity remains close to 3 for all simulations, whereas that for the horizontal velocity fluctuations is about 4 for $Pr \leq 1$ (with only a modest increase with $Pr$), and shows a rapid increase for $Pr > 1$. Thus, velocity fluctuations in the slender cell differ only slightly from Gaussian for all velocity components, as in a $\Gamma = 1$ cell using water~\citep{Qiu:PRE2001} and in a $\Gamma = 16$ cell for both low and moderate $Pr$~\citep{Pandey:APJ2021}.

\begin{figure}[t]
\begin{center}
\includegraphics[width=0.5\textwidth]{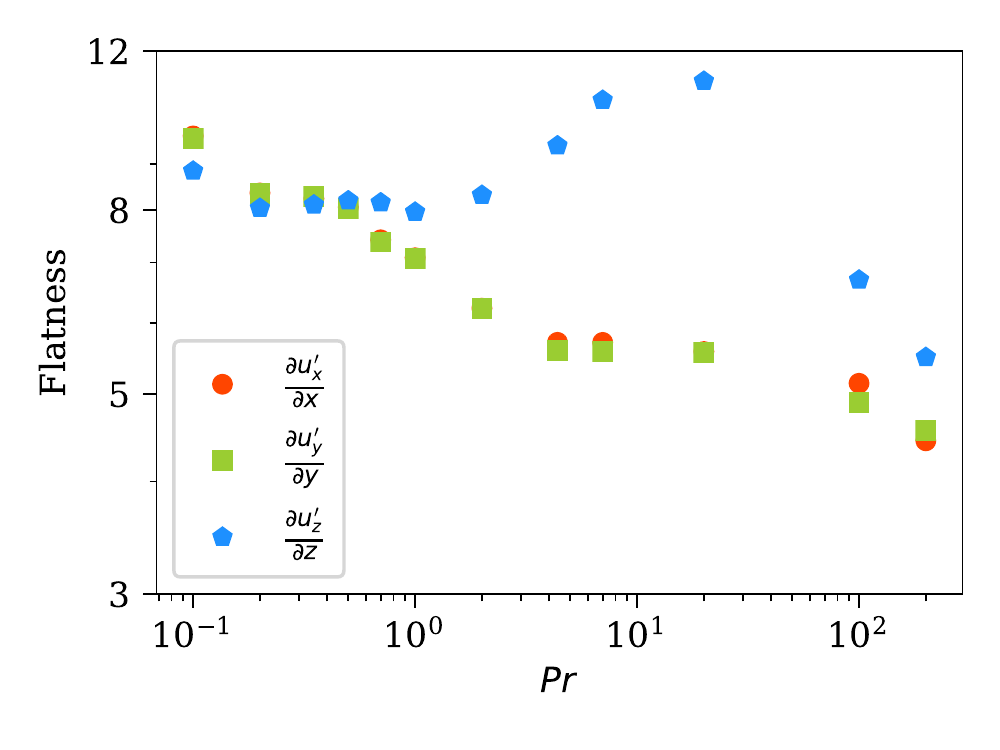}
\caption{Flatness of the horizontal velocity derivatives in the bulk region increases with decreasing $Pr$, whereas that for the vertical derivative exhibits a non-monotonic trend with $Pr$.}
\label{fig:kurtosis_der}
\end{center}
\end{figure}

Even though the velocity components are closely Gaussian, the velocity derivatives are strongly non-Gaussian, as in other turbulent flows~\citep{Sreenivasan:ARFM1997}. The PDFs of the longitudinal velocity derivatives have stretched-exponential tails, becoming increasingly spread out with decreasing $Pr$. The derivative flatness in the bulk region, plotted for various $Pr$ in Fig.~\ref{fig:kurtosis_der}, reveals an increasing trend with decreasing $Pr$. Figure~\ref{fig:kurtosis_der} also shows that the flatness for all three derivatives are similar for $Pr \leq 1$ and increase to a high value of $\approx 10$ for $Pr = 0.1$. This reaffirms that low-$Pr$ convection is characterized by an intermittent velocity field. The flatness for $\partial u_z^\prime/\partial z$ in Fig.~\ref{fig:kurtosis_der} differs significantly from that of the horizontal derivatives for moderate $Pr$ and shows a maximum at $Pr \approx 20$. We also confirm that both longitudinal derivatives in the horizontal plane give the same fourth-order moments, as expected.  

\begin{figure}[hbt]
\begin{center}
\includegraphics[width=0.5\textwidth]{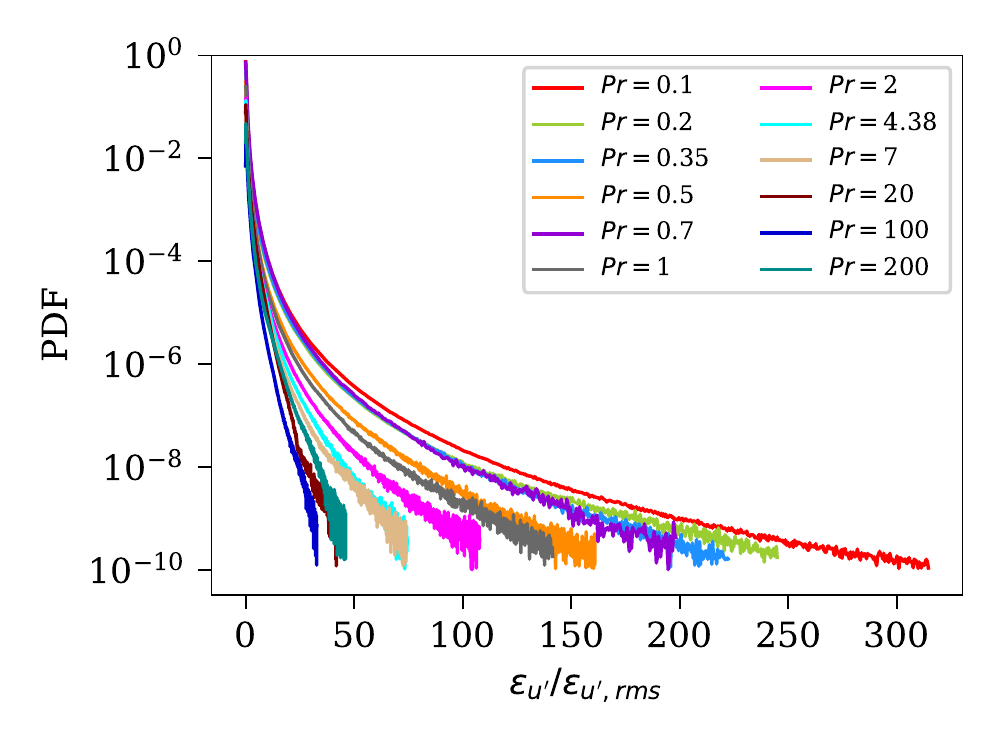}
\caption{PDFs of turbulent dissipation rate follow a stretched exponential distribution. The stretching factor increases with decreasing $Pr$, signalling increasing intermittency with decreasing $Pr$.}
\label{fig:pdf_epsv}
\end{center}
\end{figure}

The energy dissipation rate comprises all components of the velocity gradient tensor and exhibits strong intermittency in hydrodynamic as well as convective turbulence~\citep{Scheel:NJP2013, Sreenivasan:RMP1999, Ishihara:ARFM2009, Schumacher:PRE2016, Bhattacharya:POF2018}. We compute the PDFs of the turbulent viscous dissipation rate $\varepsilon_{u^\prime}$ using Eq.~(\ref{eq:eps_u}), and plot them in Fig.~\ref{fig:pdf_epsv}. Consistent with the behavior in hydrodynamic turbulence~\citep{Sreenivasan:RMP1999, Yeung:PNAS2015}, $\varepsilon_{u^\prime}$ is distributed as a stretched exponential. This behavior is true for all $Pr$, becoming increasingly so as $Pr$ decreases.

\begin{figure}
\begin{center}
\includegraphics[width=0.5\textwidth]{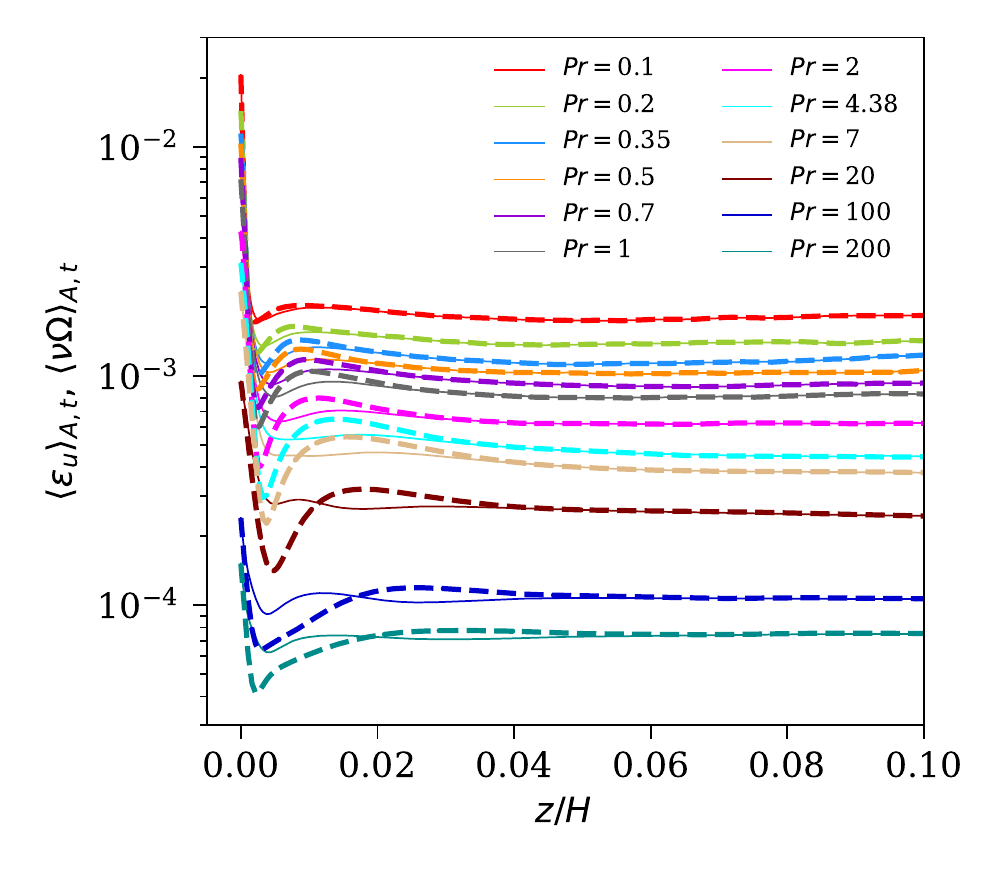}
\caption{Vertical profiles of the energy dissipation rate (solid lines) and $\nu \Omega$ (dashed lines) differ near the horizontal plate but agree perfectly in the bulk region for $z/H \geq 0.05$.}
\label{fig:epsv_z}
\end{center}
\end{figure}

The intermittency of the velocity field in a turbulent flow can also be probed by examining the vorticity field $\omega_i(x_m,t) = \epsilon_{ijk} \partial u_k/\partial x_j$, where $\epsilon_{ijk}$ is the alternating Levi-Civita tensor. The enstrophy field defined as
$\Omega(x_m,t) = \omega_j^2(x_m,t)$ quantifies the local rotation rate in the flow and its statistical properties and spatial structure closely follow those of the energy dissipation rate~\citep{Yeung:PNAS2015}. We find that the PDFs of enstrophy are also distributed as stretched exponentials, very similar to  $\varepsilon_{u^\prime}$. Moreover, in homogeneous isotropic turbulence, $\nu  \langle \Omega \rangle_V = \langle \varepsilon_u \rangle_V$---though local dissipation and enstrophy do not satisfy this equality~\citep{Yeung:PNAS2015}. The vertical profiles of $\varepsilon_u$ and $\nu \Omega$ (see Fig.~\ref{fig:epsv_z}) show that they differ from each other only near the horizontal walls. \citet{Scheel:JFM2016} also compared these profiles in a $\Gamma = 1$ cylindrical cell and found the same behavior. 


\section{Turbulent Prandtl number}
\label{sec:Prt}

The turbulent Prandtl number $Pr_t$, which is the ratio of turbulent diffusivities of momentum and thermal transports, is a flow property. In the literature, turbulent diffusivities are frequently estimated using the flux-gradient relations~\citep{Li:AR2019}: the turbulent viscosity $\nu_t$ is estimated by the ratio of the Reynolds stress $-\langle u_x^\prime u_z^\prime \rangle$ and the mean velocity gradient $\partial U_x/\partial z$, and the turbulent thermal diffusivity $\kappa_t$ by the ratio of the turbulent heat flux $-\langle u_z^\prime T^\prime \rangle$ and the mean temperature gradient $\partial \Theta/\partial z$. This approach, however, will not yield reasonable results in the bulk region of convection as the mean velocity gradient is not always well defined. In a recent work,  \citet{Pandey:PRF2021} estimated $Pr_t$ by the $k-\varepsilon$ framework (see below) and found that $Pr_t$ increases as its molecular counterpart decreases in both the Oberbeck-Boussinesq and in a specific case of non-Oberbeck-Boussinesq convection.

In this work, we estimate the turbulent viscosity and turbulent thermal diffusivity in the slender cell using the $k-\varepsilon$ approach as
\begin{eqnarray}
\nu_t = c_\nu \frac{k_u^2}{\varepsilon_{u^\prime}} \, , \\
\kappa_t = c_\kappa \frac{k_u k_T}{\varepsilon_{T^\prime}}, 
\end{eqnarray}
where $k_u = u_i^{\prime 2}/2$ and $k_T = T^{\prime 2}$ are the turbulent kinetic energy and thermal variance, respectively. The turbulent thermal dissipation rate is computed using Eq.~(\ref{eq:eps_T}) from the fluctuating temperature field. Since we are primarily interested in the variation of $Pr_t$ with $Pr$, we treat the coefficients $c_\nu$ and $c_\kappa$ as constants independent of $Ra$ and $Pr$. Note that this is an assumption. Our choice of the coefficient $c_\nu = 0.09$ is the same as used in turbulence models~\citep{Davidson:book2004, Yakhot:IJHMT1987, Yakhot:JSC1986}.

\begin{figure}[hbt]
\begin{center}
\includegraphics[width=0.5\textwidth]{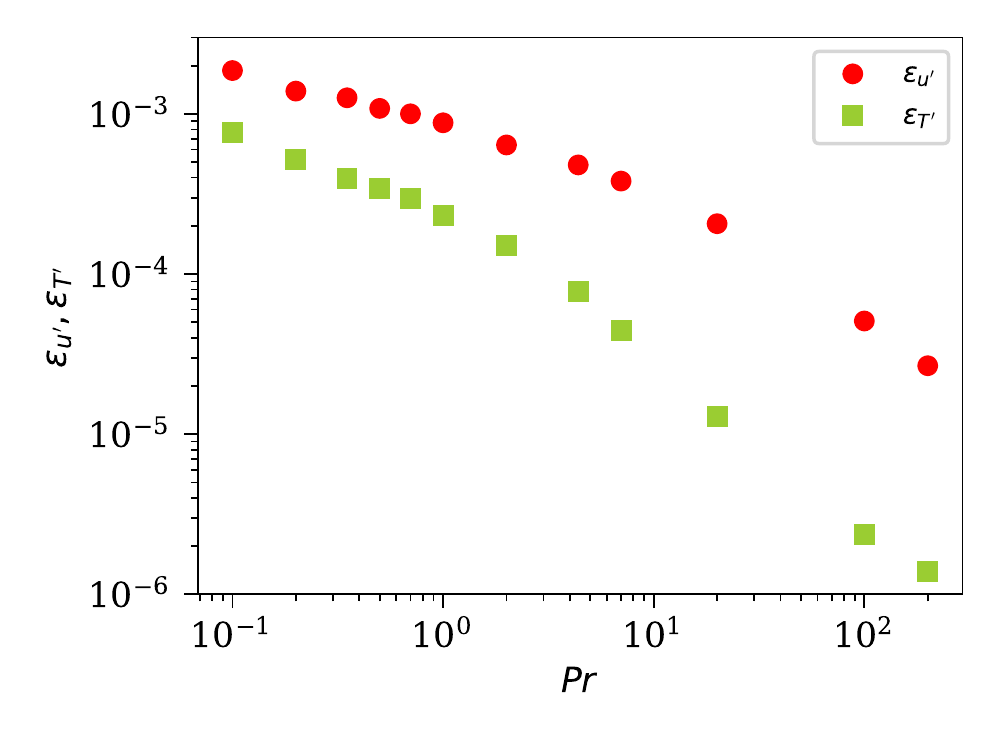}
\caption{Turbulent energy and thermal dissipation rates in the bulk region, $0.2H \leq z \leq 0.8H$, decrease gradually for low Prandtl numbers, but relatively steeply in the viscous regions at high Prandtl numbers.}
\label{fig:diss_Pr}
\end{center}
\end{figure}

To see the variation of the turbulent diffusivities in the vertical direction, we compute their depth profiles using $k_u(z)$, $k_T(z)$, $\varepsilon_{u^\prime}(z)$, and $\varepsilon_{T^\prime}(z)$. We observe that horizontally-averaged turbulent kinetic energy and thermal variance vanish at the horizontal plates due to the imposed boundary conditions, but grow rapidly as we move from the horizontal boundaries. On the other hand, the averaged dissipation rates are highest at the plates as the strongest gradients are observed in the vicinity of the plates, and decrease rapidly as one moves farther away from them. However, all these quantities are nearly unchanging in the bulk region~\citep{Pandey:PRF2021}. We show the bulk-averaged turbulent dissipation rates as a function of $Pr$ in Fig.~\ref{fig:diss_Pr} and observe that both $\varepsilon_{u^\prime}$ and $\varepsilon_{T^\prime}$ decrease with $Pr$~\citep{Bhattacharya:PRF2021}. This is expected from the exact relations, which yield $\varepsilon_u \sim (Nu-1)/\sqrt{RaPr}$ and $\varepsilon_T \sim Nu/\sqrt{RaPr}$; as $Nu$ scales slower than $Pr^{0.5}$~\citep{Pandey:EPL2021}, both $\varepsilon_{u^\prime}$ and $\varepsilon_{T^\prime}$ decrease with $Pr$. Furthermore, the rapid decrease of the turbulent dissipation rates at high Prandtl numbers in Fig.~\ref{fig:diss_Pr} is again consistent with the scaling $Nu(Pr)$ for $Ra = 3 \times 10^{10}$ discussed in \citet{Pandey:EPL2021}.

\begin{figure}[H]
\begin{center}
\includegraphics[width=0.5\textwidth]{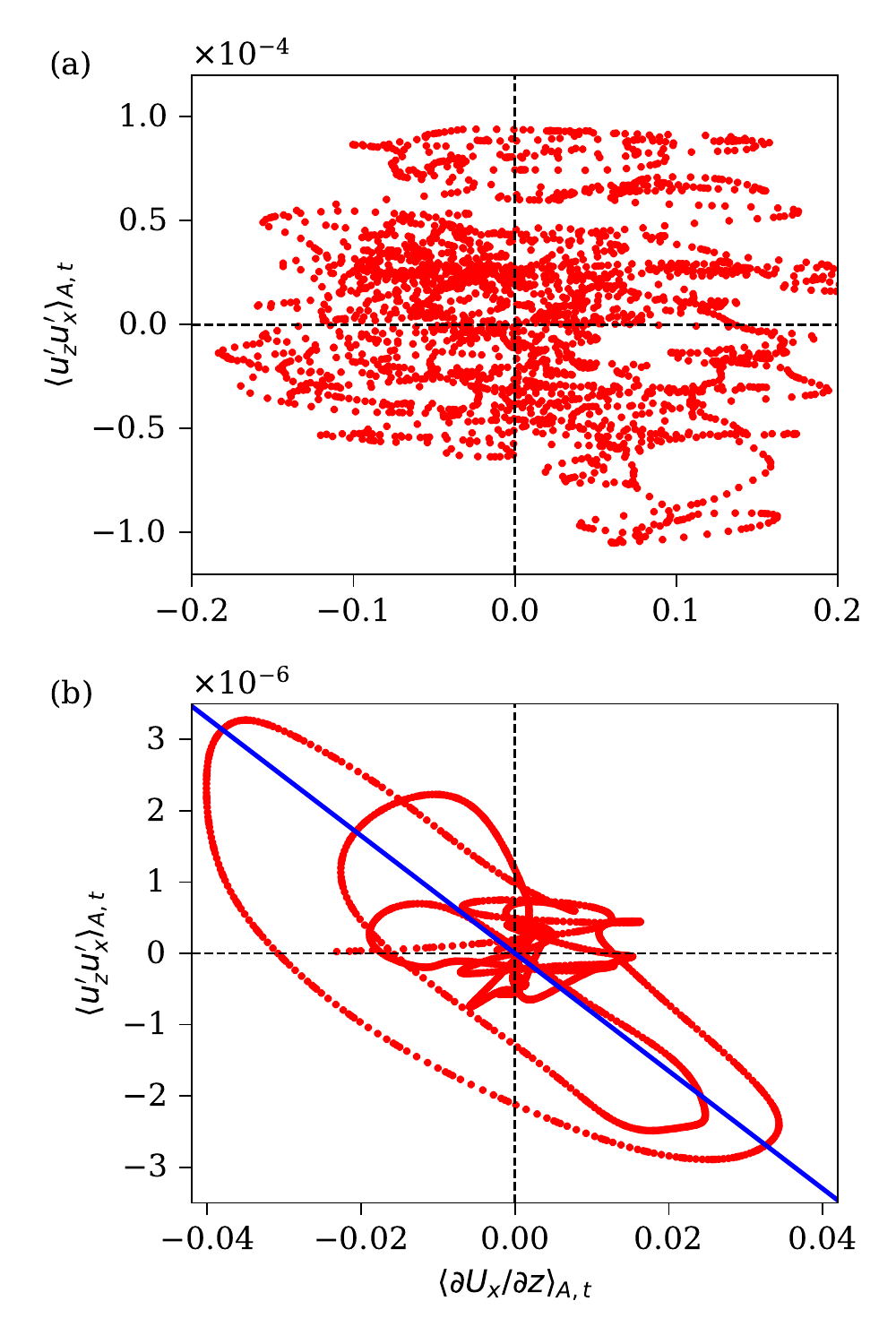}
\caption{Turbulent momentum flux $\langle u_z^\prime u_x^\prime \rangle_{A,t}(z)$ as a function of the mean velocity gradient $\langle \partial U_x/\partial z \rangle_{A,t}(z)$ for slender cell simulations at (a) $Pr = 0.35$ and (b) $Pr = 100$, $\Gamma = 0.1$. Strongly fluctuating velocity field in low-$Pr$ convection results in a nearly homogeneous distribution of the phase points, which makes it challenging to estimate the turbulent viscosity using the flux-gradient method. The straight line in (b) can be thought to represent the state around which the data are organized but this interpretation cannot be imagined for (a).}
\label{fig:phase_nu}
\end{center}
\end{figure}

\begin{figure}[H]
\begin{center}
\includegraphics[width=0.5\textwidth]{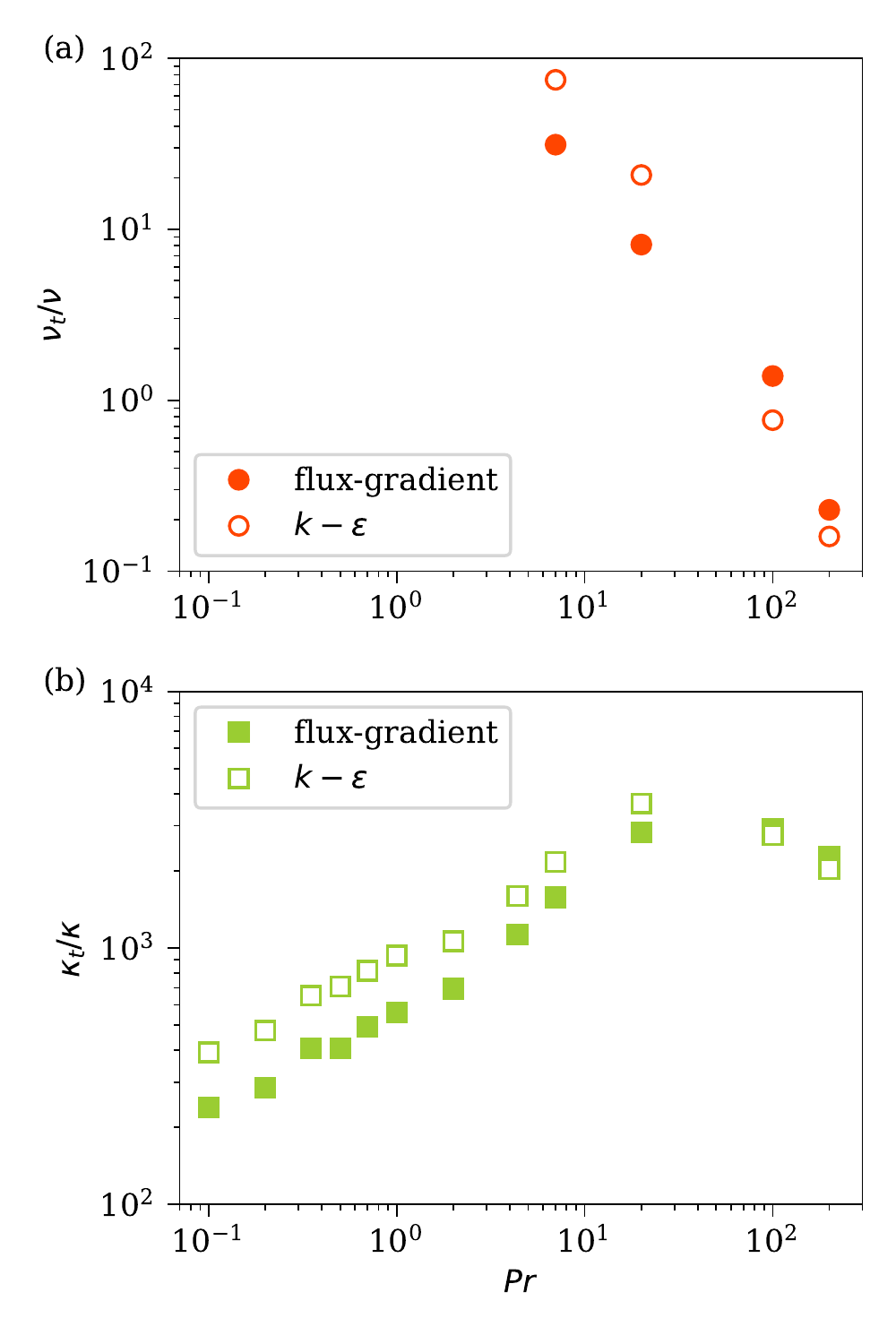}
\caption{Ratios of the turbulent and molecular (a) viscosities and (b) thermal diffusivities in the bulk region from two approaches agree reasonably with each other and exhibit closely similar trends with $Pr$.}
\label{fig:turb_diffs_compare}
\end{center}
\end{figure}

As mentioned earlier, estimating turbulent viscosity using the flux-gradient approach is challenging in RBC. To demonstrate this, we show a phase plot of the turbulent momentum flux $\langle u_z^\prime u_x^\prime \rangle_{A,t}(z)$ as a function of the mean velocity gradient $\langle \partial U_x/\partial z \rangle_{A,t}(z)$ in Fig.~\ref{fig:phase_nu} for $Pr = 0.35$ (panel a) and $Pr = 100$ (panel b), $\Gamma = 0.1$. The phase points are distributed homogeneously in all four quadrants for the low-$Pr$ case, whereas the distribution for high-$Pr$ case can be thought to be close to the diagonal line shown. Distributions similar to those in Fig.~\ref{fig:phase_nu}(a) and Fig.~\ref{fig:phase_nu}(b) are observed for all $Pr \leq 4.38$ and $Pr \geq 7$, respectively. Thus, the turbulent viscosity 
\begin{equation}
\nu_t(z) = -\frac{\langle u_z^\prime u_x^\prime \rangle_{A,t}(z)}{ \langle \partial U_x/\partial z \rangle_{A,t}(z)}
\label{rans}
\end{equation}
frequently alters sign in low-$Pr$ flows as the altitude varies, which poses difficulty in estimating $\nu_t$. The same holds true when $x$ in Eq.~\eqref{rans} is replaced by $y$. The absence of a coherent large-scale flow with a properly defined mean shear rate, in particular for the low Prandtl numbers, prohibits the application of the flux-gradient scheme, see also \cite{Schindler:PRL2022} for recent experiments in liquid metals. For $Pr \geq 7$, we estimate $\nu_t$ as the slope of the line and compare it with that estimated using the $k-\varepsilon$ method as $\nu_t = k_u^2/\varepsilon_{u^\prime}$. The ratio of the turbulent and molecular viscosities $\nu_t/\nu$ from the two methods are plotted as a function of $Pr$ in Fig.~\ref{fig:turb_diffs_compare}(a). The agreement is reasonable and, similar to the findings of \citet{Pandey:PRF2021}, $\nu_t/\nu$ decreases with increasing $Pr$.

A slight non-zero mean temperature gradient is indeed found for the slender cell~\citep{Iyer:PNAS2020, Pandey:EPL2021}, see Fig.~\ref{fig:T_z}. This enables us to estimate $\kappa_t$ using the flux-gradient approach; we find that the turbulent heat flux $\langle u_z^\prime T^\prime \rangle_{A,t}(z)$ remains positive in the bulk region of the slender cell. We compute the turbulent diffusivity in the bulk region, i.e. for $0.2H \leq z \leq 0.8H$, as 
\begin{equation}
\kappa_t = -\frac{ \langle u_z^\prime T^\prime \rangle_\mathrm{bulk}}{\langle \partial \Theta/\partial z \rangle_\mathrm{bulk}}
\end{equation}
and plot $\kappa_t/\kappa$ as a function of $Pr$ in Fig.~\ref{fig:turb_diffs_compare}(b). For comparison, we also show the diffusivity computed from the $k-\varepsilon$ method as $\kappa_t = k_u k_T/\varepsilon_{T^\prime}$; both schemes yield similar results. The increasing trend of $\kappa_t/\kappa$ up to $Pr = 20$ is again consistent with the findings of \citet{Pandey:PRF2021}.

\begin{figure}[H]
\begin{center}
\includegraphics[width=0.5\textwidth]{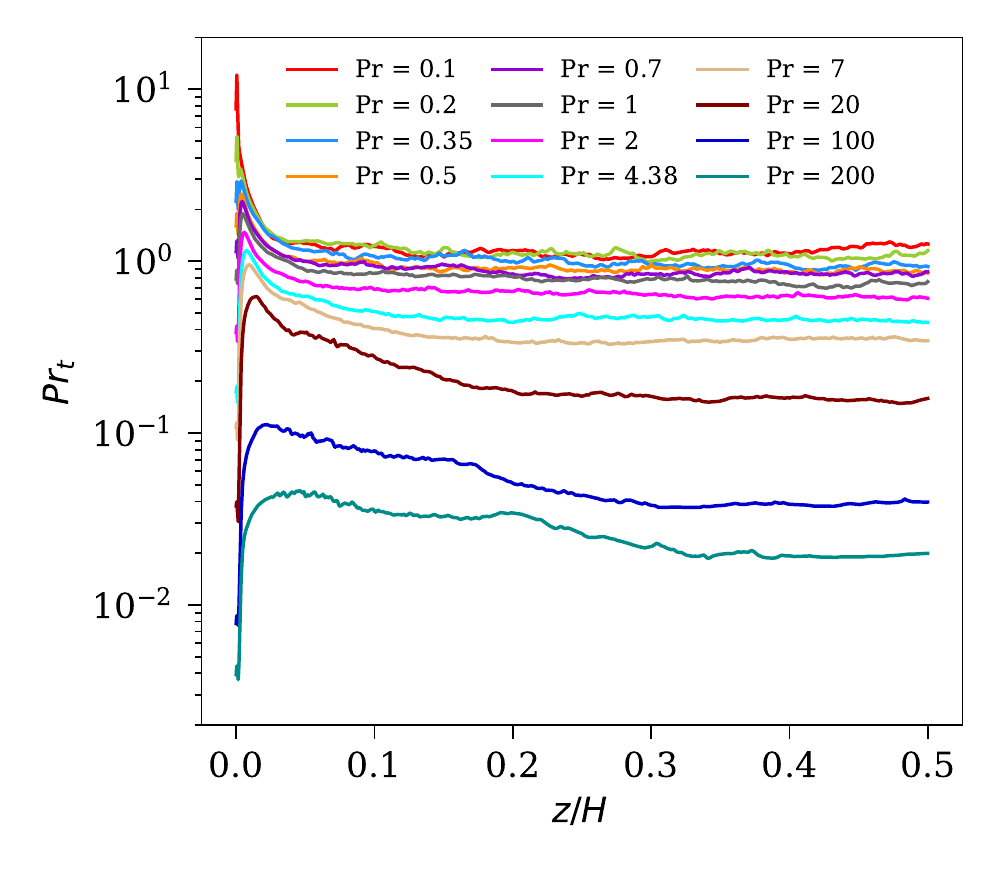}
\caption{Turbulent Prandtl number averaged over the top and bottom halves of the slender cell varies sharply near the wall but remains nearly unvarying in the bulk region. The corresponding bulk-averaged $Pr_t$ are shown in Fig.~\ref{fig:Prt_Pr} as open diamonds.}
\label{fig:Prt_z}
\end{center}
\end{figure}

We compute the depth-dependent turbulent Prandtl number as
\begin{equation}
Pr_t(z) = \frac{\nu_t(z)} {\kappa_t(z)} = \frac{c_\nu}{c_\kappa} \frac{k_u(z)}{k_T(z)} \frac{\varepsilon_{T^\prime}(z)}{  \varepsilon_{u^\prime}(z)}
\label{Prturb}
\end{equation}
and show it in Fig.~\ref{fig:Prt_z}. Again, the profiles are averaged over upper and lower halves of the domain. As discussed in \citet{Pandey:PRF2021}, we use $c_\nu/c_\kappa = 1.43$ to obtain $Pr_t \approx 0.85$ for $Pr = 0.7$. Note that $Pr_t$ remains small in the near-wall region. This is reasonable as the turbulent fluctuations are weaker near the plates, leading to insignificant turbulent heat and momentum transports compared to those in the bulk region. Figure~\ref{fig:Prt_z} also shows that beyond a rapid increase in the near-wall region, $Pr_t(z)$ relaxes to a nearly unchanging value in the bulk, $z \geq 0.2H$.

\begin{figure}[H]
\begin{center}
\includegraphics[width=0.5\textwidth]{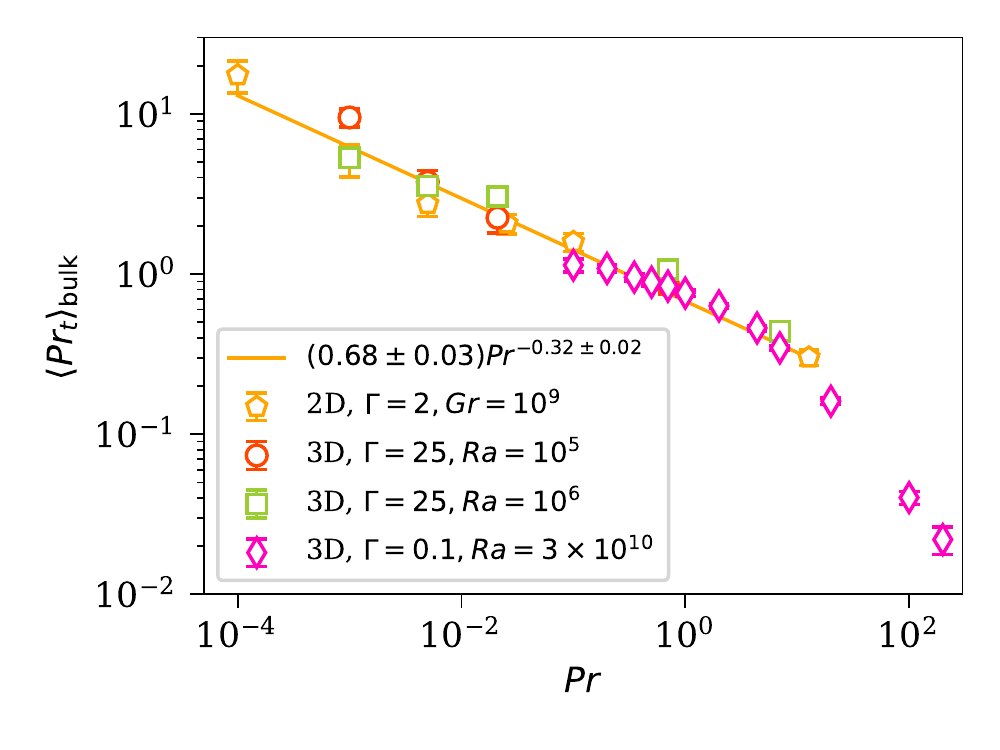}
\caption{Turbulent Prandtl number increases with decreasing molecular $Pr$. The error bars indicate the standard deviation of $Pr_t$ in the bulk region. Data from the slender cell collapse well on those obtained from the extended domain as well as from  a wider two-dimensional domain.}
\label{fig:Prt_Pr}
\end{center}
\end{figure}

We now compute the averaged turbulent Prandtl number in the bulk region and plot it as a function of $Pr$ in Fig.~\ref{fig:Prt_Pr}. We see that $\langle Pr_t \rangle_\mathrm{bulk}$ increases with decreasing $Pr$. For comparison, we also show $\langle Pr_t \rangle_\mathrm{bulk}$ for convection in a two-dimensional (2D) box of $\Gamma = 2$ (from \citet{Pandey:PRF2021}) and for convection in the $\Gamma = 25$ square cell (from \citet{Pandey:JFM2022}). It is clear from Fig.~\ref{fig:Prt_Pr} that the variation of $Pr_t$ with $Pr$ is very similar for all three cases for $Pr \lesssim 20$. The scaling $\langle Pr_t \rangle_\mathrm{bulk} \sim Pr^{-1/3}$, which was obtained from the 2D data~\citep{Pandey:PRF2021}, seems to describe the turbulent Prandtl number variation in both the slender and extended domains, for $Pr \lesssim 20$. For $Pr > 20$, the slender cell data reveals that $\langle Pr_t \rangle_\mathrm{bulk}$ falls much faster with $Pr$. Our findings suggest that the turbulent Prandtl number, as obtained here, is not sensitive to the aspect ratio. We stress that the plot collects results covering six orders of magnitude in the molecular Prandtl number. 

Finally, we provide an estimate of $Pr_t$ using the exact relations~\citep{Shraiman:PRA1990} and the observed scaling relations of the velocity and temperature fluctuations in RBC. The exact relations~\cite{Shraiman:PRA1990} are
\begin{align}
\varepsilon_T & =  \kappa \frac{(\Delta T)^2}{H^2} Nu \, , \\
\varepsilon_u & =  \frac{\nu^3}{H^4} \frac{(Nu-1) Ra}{Pr^2}
\end{align}
and the scaling relations for the turbulent kinetic and thermal energies are 
\begin{align}
k_u & = u_i^{\prime 2}/2 \sim u_\mathrm{rms}^2 \, \\
k_T & = T^{\prime 2} \sim T_\mathrm{rms}^{2}.
\end{align}
We recall that the primes indicate local root-mean-square values while the suffix ``rms" indicates global root-mean-square values; our data show that $T^\prime_\mathrm{rms}$ is smaller than $T_\mathrm{rms}$ but the two have nearly the same $Pr$-dependence.  The turbulent thermal dissipation rate $\varepsilon_{T^\prime}$ could be considered to scale the same way as the total turbulent energy dissipation, $\varepsilon_T$, and the factor $Nu-1$ could be easily replaced by $Nu$. We further use the observed relation $T_\mathrm{rms} \sim Pr^{-\zeta} \Delta T$. Incorporating these assumptions, we obtain the following relations:
\begin{align}
k_u & \sim  u_\mathrm{rms}^2 \sim Re^2 \frac{\nu^2}{H^2} \, , \\
k_T & \sim  T_\mathrm{rms}^2 \sim Pr^{-2\zeta} (\Delta T)^2 \, , \\
\varepsilon_{T^\prime} & \sim  \kappa \frac{(\Delta T)^2}{H^2} Nu \, , \\
\varepsilon_{u^\prime} & \sim  \frac{\nu^3}{H^4} \frac{Nu Ra}{Pr^2} \, ,
\end{align}
which yield the turbulent Prandtl number \eqref{Prturb} to be
\begin{equation}
Pr_t \sim Re^2 Pr^{1+2\zeta} Ra^{-1}.  \label{eq:Pr_t_2}
\end{equation}
We observe $Re \sim Pr^{-2/3}Ra^{1/2}$ for low and moderate $Pr$~\citep{Pandey:EPL2021, Scheel:PRF2017}, and $Re \sim Pr^{-1} Ra^{3/5}$ for high $Pr$~\citep{Silano:JFM2010, Pandey:PRE2014, Pandey:EPL2021}. Further, from Fig.~\ref{fig:T_rms} we obtain $\zeta \approx 0.17$ for $Pr \leq 2$ and $\zeta \approx 0.36$ for $Pr \geq 2$. Using these exponent values in Eq.~\eqref{eq:Pr_t_2} results in $Pr_t \sim Pr^{0}$ for low-$Pr$, and $Pr_t \sim Pr^{-0.28}$ for moderate-$Pr$, which differ from our findings.

In reality, in analogy to DNS results in homogeneous and isotropic turbulence~\citep{Donzis:JFM2005}, it is quite conceivable that the relations for both dissipation rates in the bulk follow
\begin{align}
\varepsilon_{T^\prime} &  = \gamma(Re,Pr) \frac{u_\mathrm{rms} T_\mathrm{rms}^2}{\ell}  \, , \\
\varepsilon_{u^\prime} & =  \beta(Re,Pr) \frac{u_\mathrm{rms}^3}{\ell} \, .
\end{align}
Equation~\eqref{Prturb} would then lead to
\begin{equation}
Pr_t \approx \frac{\gamma(Re,Pr)}{\beta(Re,Pr)}\,.
\end{equation}
Reference~\citep{Pandey:JFM2022} provided an analysis of $\beta(Re)$ in large aspect ratio cells, and showed that the resulting dependence on $Pr$ of the turbulent Prandtl number is similar to that in Fig.~\ref{fig:Prt_Pr}. But the data and analysis of this point are limited at present and it is fair to say that we cannot draw firm conclusions from such theoretical arguments; we leave this point open for future work.
\section{Conclusions}
\label{sec:conclu}
The goal of this work has been to compare direct numerical simulations on convection in a slender cell, over a wide range of Prandtl numbers, with those in large aspect ratio domains. Further, we collect data on the dependence of turbulent Prandtl number on the (molecular) Prandtl number \cite{Pandey:PRF2021,Pandey:JFM2022}.

We studied the characteristics of the temperature and velocity fluctuations in a slender cylinder of aspect ratio 0.1 for a fixed and relatively high Rayleigh number $Ra=3\times 10^{10}$, with $Pr$ varying over three orders of magnitude. We compared the fluctuations in the slender cell with those in a horizontally-extended square box of $\Gamma = 25$ and found similarities as well as some differences. As examples of similarities, the root-mean-square temperature fluctuations scale similarly with $Pr$ in both the configurations. The $Pr$-dependence of the root-mean-square velocity fluctuations is also similar.

Another important point of similarity is that the turbulent Prandtl number behaves very similarly in both the convection cells in the overlapping range of $Pr$. The summary of all data shown of Fig. \ref{fig:Prt_Pr}, covering six orders of magnitude in molecular Prandtl number, is that the systematically decaying trend of $\langle Pr_t\rangle_{\rm bulk} (Pr)$ is independent of whether the convection is 2D or 3D, as well as the aspect ratio of the Rayleigh-B\'{e}nard cell (within the range covered here). We stress here again that, in contrast to turbulent shear flows, a mean large-scale shear in an extended domain will be zero in the bulk region so that the flux-gradient Boussinesq model cannot be applied globally. We observed the same for the lowest Prandtl numbers in a our setup. For higher $Pr$, the $Pr_t$ values deduced from both methods were effectively the same for the low aspect ratio case.

Yet another point of similarity is that the probability density functions of the temperature and velocity fluctuations show very similar $Pr$-dependencies for low and high aspect ratios. Specifically, the PDFs become strongly non-Gaussian at high Prandtl numbers, whereas they are very close to Gaussian in low-$Pr$ convection. The vertical velocity shows similar Gaussian distribution for both cases.

However, there are some differences. For example, the distribution of the horizontal velocity components is Gaussian in the extended flow, whereas discernible departures are observed in the slender cell near the wall. We can thus conclude that even though the slender cell geometry constrains turbulent convection, its many statistical results are the same as those for larger domains. It thus provides an appropriate testing bed to advance to even higher Rayleigh numbers than the present limit of $Ra=10^{15}$ \cite{Iyer:PNAS2020}.   

\section*{Acknowledgement}
We are honored to contribute to this special issue for Charlie Doering, who inspired our research via fruitful discussions over several decades; he was a positive force and will be missed. KRS, JS and AP are dismayed to note the untimely demise on July 24 of their coauthor, Ravi Samtaney, during the revision of this manuscript. JS wishes to thank the Isaac Newton Institute for Mathematical Sciences, Cambridge, for support and hospitality during the program {\it Mathematical aspects of turbulence: where do we stand?} where part of this work was undertaken. This research was supported by the KAUST Office of Sponsored Research under Award URF/1/4342-01, and also by EPSRC grant EP/R014604/1. The three-dimensional large-aspect-ratio data were obtained at the SuperMUC-NG compute cluster within the project pn68ni of the Leibniz Rechenzentrum Garching. The authors gratefully acknowledge {\sc Shaheen II} of King Abdullah University of Science and Technology, Saudi Arabia (under Project No. k1491), as well as {\sc Dalma} and {\sc Jubail} clusters at NYU Abu Dhabi for providing computational resources via the Center for Space Science (grant G1502). 



\end{document}